\documentstyle[12pt]{article}
\newcommand{\be}{\begin{equation}}
\newcommand{\ee}{\end{equation}}
\newcommand{\bea}{\begin{eqnarray}}
\newcommand{\eea}{\end{eqnarray}}
\newcommand{\beaa}{\begin{eqnarray*}}
\newcommand{\eeaa}{\end{eqnarray*}}

\newcommand{\BB}{{{\rm I} \kern -2pt \rlap {\rm B} \kern +8pt}}

\advance\textheight by \topskip
\makeatletter

\@addtoreset{equation}{section}
\def\section{\@startsection {section}{1}{\z@}{-3.5ex plus -1ex minus
 -.2ex}{2.3ex plus .2ex}{\large\bf\centering}}
\def\subsection{\@startsection{subsection}{2}{\z@}{-3.25ex plus -1ex minus -.2ex}{1.5ex plus .2ex}{\bf}}
\def\subsubsection{\@startsection{subsubsection}{3}{\z@}{-3.25ex plus -1ex minus -.2ex}{1.5ex plus .2ex}{\sl}}
\makeatother

\begin{document}

\baselineskip 18pt \parindent 12pt \parskip 10pt

\begin{titlepage}

\begin{center}
{\Large {\bf On linearization of super sine-Gordon
equation}}\\\vspace{1.5in} {\large M.
Siddiq\footnote{%
On leave of absence from PRD, PINSTECH, Nilore Islamabad
(Pakistan).} and M. Hassan }\vspace{0.15in}

{\small{\it Department of Physics,\\ University of the Punjab,\\
Quaid-e-Azam Campus,\\ Lahore-54590, Pakistan.}}

\end{center}

\vspace{1cm}
\begin{abstract}
Two sets of super Riccati equations are presented which result in
two linear problems of super sine-Gordon equation. The linear
problems are then shown to be
 related to each other by a super gauge transformation and to the super B\"{a}cklund
transformation of the equation.
\end{abstract}
\vspace{1cm} PACS: 02.30.Ik
\end{titlepage}

\section{Introduction}

\bigskip During the last few decades, much progress has been made in the
study of nonlinear evolution equations with soliton solutions.
Some of these equations correspond to integrable models of field
theory and are solvable by the inverse scattering method in the
scheme of Zakharov-Shabat (ZS) and Ablowitz, Kaup, Newell and
Segur (AKNS) \cite{ref1}-\cite{lax}. The inverse scattering method
and ZS/AKNS scheme are the general procedures which associate a
linear system with the nonlinear evolution equations and lead to
soliton solutions. This scheme incorporates all the well known
integrable models such as the Korteveg deVries equation (KdV),
modified KdV (mKdV) equation, Liouville theory, sine-Gordon
theory, Schrodinger equation, sigma model etc. The associated
linear system is related to the existence of infinitely many local
conserved quantities in involution, which guarantee the the
classical and quantum integrability of the model. Moreover, the
B\"{a}cklund transformation and its Riccati form are one of the
direct methods of obtaining new soliton solutions and local
conserved quantities of a given integrable model. The B\"{a}cklund
transformation can be derived by a suitable transformation from
the Riccati equations \cite{chen} and through these Riccati
equations the B\"{a}cklund transformation is related to the
inverse scattering method and the ZS/AKNS scheme. This suggests
that the Riccati equations related to the inverse scattering
method can be used to derive an infinite number of local conserved
quantities.

More recently, many techniques have been developed to study the
supersymmetric generalizations of integrable models, for example the
supersymmetric extension of inverse scattering method, ZS/AKNS scheme, the
Lax formalism, zero curvature formalism, conservation laws etc \cite{Kuper}-%
\cite{usman}. Among the several super integrable models, perhaps the super
sine-Gordon equation is the physically most interesting equation. The super
sine-Gordon equation is integrable as it contains infinitely many local
conserved quantities \cite{ferrara} which survive quantization, rendering
the theory quantum integrable and making it possible to compute the exact
S-matrices \cite{ber}-\cite{ahn 91}. The linear representation of the super
sine-Gordon equation has been proposed in \cite{Sc1} and using that a
rigorous proof of the existence of infinitely many local conserved
quantities has been derived. So far there exists no systematic way of
establishing a linear representation of super sine-Gordon equation which
could be manifestly related to the B\"{a}cklund transformation and its
Riccati form. The purpose of this work is to present a systematic approach
to determine the desired linear representation of the super sine-Gordon
equation.

\bigskip In our analysis, we start with two sets of Riccati equations whose
compatibility condition is the super sine-Gordon equation. The two
sets of Riccati equations are shown to be related to the
B\"{a}cklund transformation of the super sine-Gordon equation. The
linearization of Riccati equations yields two linear systems
associated with the super sine-Gordon equation. Both the linear
systems are shown to be equivalent, related to each other by a
super gauge transformation.

We follow the general procedure of writing manifestly supersymmetric
sine-Gordon equation. The equation is defined in two dimensional
super-Minkowski space with bosonic light-cone coordinates $x^{\pm }$ and
fermionic coordinates $\theta ^{\pm },$ which are Majorana spinors. The
covariant superspace derivatives are defined by
\[
D_{\pm }=\frac{\partial }{\partial \theta ^{\pm }}-i\theta ^{\pm }\partial
_{\pm },\,\,\,\,\,\,\,D_{\pm }^{2}=-i\partial _{\pm
},\,\,\,\,\,\,\,\,\,\{D_{+},D_{-}\}=0,
\]
where $\{\,,\,\}$ is an anti-commutator. The superspace lagrangian density
of $N=1$ super sine-Gordon theory is
\begin{equation}
{\cal L}(\Phi )=\frac{i}{2}D_{+}\Phi D_{-}\Phi +\cos \Phi ,  \label{lag1}
\end{equation}
where $\Phi $ is a real scalar superfield. The superfield evolution equation
follows from the lagrangian and is given by
\begin{equation}
D_{+}D_{-}\Phi =i\sin \Phi .  \label{sin}
\end{equation}
The equation (\ref{sin}) is invarient under $N=1$ supersymmetry
transformations.

\section{{{Super Riccati equations and super B\"{a}cklund transformation}}}

The super sine-Gordon equation appears as the compatibility condition of the
following set of Riccati equations
\begin{eqnarray}
D_{+}{\cal N} &=&\frac{1}{2}\sqrt{\frac{2}{\lambda }}g\left[ {\cal N}\,\cos
\Phi +{\cal N}^{2}\sin \Phi \right] ,  \label{ric1} \\
D_{-}{\cal N} &=&\frac{\sqrt{2\lambda }}{2}g{\cal N}+(1+{\cal N}%
^{2})\,D_{-}\Phi ,  \nonumber
\end{eqnarray}
together with
\begin{eqnarray}
{\cal N}D_{+}g &=&-i\sqrt{\frac{2}{\lambda }}\left[ -{\cal N}\cos \Phi +\sin
\Phi \right] ,  \label{ric2} \\
{\cal N}D_{-}g &=&-i{\cal N}\sqrt{2\lambda }+gD_{-}\Phi ,  \nonumber
\end{eqnarray}
where ${\cal N}$ and $g$ are scalar and spinor superfields, respectively,
and $\lambda $ is the spectral parameter. The fermionic superfield $g$ is
introduced because of supersymmetry and the oddness of the superspace
derivatives $D_{\pm }$. The compatibility condition of equations (\ref{ric1}%
) and (\ref{ric2}) is equation (\ref{sin}). In the limit when fermions are
set equal to zero, the super Riccati system reduces to pure bosonic Riccati
system of the sine-Gordon equation.

The super sine-Gordon equation can also be expressed as the compatibility
condition of another set of Riccati equatios which is
\begin{eqnarray}
D_{+}\Gamma &=&\sqrt{\frac{2}{\lambda }}\frac{g}{4}\left[ (\Gamma -1)\exp
(i\Phi )+\Gamma (\Gamma -1)\exp (-i\Phi )\right] ,  \label{ric3} \\
D_{-}\Gamma &=&\sqrt{2\lambda }\frac{g}{4}\left( \Gamma ^{2}-1\right)
+2i\Gamma D_{-}\Phi ,  \nonumber
\end{eqnarray}
together with
\begin{eqnarray}
\left( \Gamma -1\right) D_{+}g &=&-i\sqrt{\frac{2}{\lambda }}\left[ \exp
(i\Phi )-\Gamma \exp (-i\Phi )\right] ,  \label{ric4} \\
\left( \Gamma -1\right) D_{-}g &=&-i\sqrt{2\lambda }\left( \Gamma -1\right)
+i\left( \Gamma +1\right) gD_{-}\Phi ,  \nonumber
\end{eqnarray}
where $\Gamma $ is a scalar superfield. In what follows, we shall
see that the linearization of set of Riccati equations
(\ref{ric1})-(\ref{ric2}) yields the linear system of
\cite{Sc}-\cite{Sc1}, while the other set (\ref
{ric3})-(\ref{ric4}) gives its gauge equivalent form.

The given two sets of Riccati equations (\ref{ric1})-(\ref{ric2})
and (\ref {ric3})-(\ref{ric4}) can also be used to derive the
B\"{a}cklund transformation of the super sine-Gordon equation. In
order to do so, we make transformations on superfields ${\cal N}$ and $\Gamma $ as ${\cal N}$%
{\cal \,}$\rightarrow \tan \left( \frac{\Phi +\tilde{\Phi}}{2}\right) $ and $%
\Gamma \rightarrow \exp i(\Phi +\tilde{\Phi}),$ and substitute
them in (\ref {ric1})-(\ref{ric2}) and (\ref{ric3})-(\ref{ric4}),
respectively. This yields
\begin{eqnarray}
D_{+}(\Phi +\tilde{\Phi}) &=&\sqrt{\frac{2}{\lambda }}\,\,g\,\,\sin \left(
\frac{\Phi +\tilde{\Phi}}{2}\right) \,\cos \left( \frac{\Phi -\tilde{\Phi}}{2%
}\right) ,  \label{bt1} \\
D_{-}(\Phi -\tilde{\Phi}) &=&-\sqrt{2\lambda }\,\,g\,\sin \left( \frac{\Phi +%
\tilde{\Phi}}{2}\right) \,\,\cos \left( \frac{\Phi +\tilde{\Phi}}{2}\right) ,
\nonumber \\
D_{+}g &=&-i\sqrt{\frac{2}{\lambda }}\sin \left( \frac{\Phi -\tilde{\Phi}}{2}%
\right) \,\csc \left( \frac{\Phi +\tilde{\Phi}}{2}\right) ,  \nonumber \\
D_{-}g &=&-i\sqrt{2\lambda }+\cot \left( \frac{\Phi +\tilde{\Phi}}{2}\right)
g\,D_{-}\Phi ,  \nonumber
\end{eqnarray}
which is the super B\"{a}cklund transformation for the super
sine-Gordon equation. The compatibility condition of the
B\"{a}cklund transformation (\ref{bt1}) is the super sine-Gordon
equation for both $\Phi $ and $\tilde{\Phi}$ separately. One can
also relate the transformation (\ref{bt1})
to the known super B\"{a}cklund transformation of \cite{Ku}, by choosing $%
\,g=f\,\,\csc \left( \frac{\Phi +\tilde{\Phi}}{2}\right) ,$ to get
\begin{eqnarray}
D_{+}(\Phi +\tilde{\Phi}) &=&\sqrt{\frac{2}{\lambda }}\,\,f\,\,\cos \left(
\frac{\Phi -\tilde{\Phi}}{2}\right) ,  \label{bt2} \\
D_{-}(\Phi -\tilde{\Phi}) &=&-\sqrt{2\lambda }\,\,f\,\,\cos \left( \frac{%
\Phi +\tilde{\Phi}}{2}\right) ,  \nonumber \\
D_{+}f &=&-i\sqrt{\frac{2}{\lambda }}\sin \left( \frac{\Phi -\tilde{\Phi}}{2}%
\right) ,  \nonumber \\
D_{-}f &=&-i\sqrt{2\lambda }\sin \left( \frac{\Phi +\tilde{\Phi}}{2}\right) .
\nonumber
\end{eqnarray}
The super B\"{a}cklund transformations (\ref{bt1}) and (\ref{bt2})
reduce to the B\"{a}cklund transformation of the purely bosonic
sine-Gordon equation when fermions are set equal to zero. In ref.
\cite{Ku} the B\"{a}cklund transformation (\ref{bt2}) has been
used to derive an infinite number of conservation laws of the
super sine-Gordon equation.

\section{Linear systems and Lax representation in superspace}

The two sets of Riccati equations (\ref{ric1})-(\ref{ric2}) and (\ref{ric3}%
)-(\ref{ric4}) can be transformed to systems of linear equations for bosonic
superfields $\Psi _{1}$, $\Psi _{2}$ and a fermionic superfield $\Psi _{3}$
for the set (\ref{ric1})-(\ref{ric2}) by change of variables ${\cal N}=\frac{%
\Psi _{1}}{\Psi _{2}}\,$, $\,g=\frac{\Psi _{3}}{\Psi _{1}}\,$ and also for
bosonic superfields $\Phi _{1},\Phi _{2}$ and $\ $a fermionic superfield $%
\Phi _{3}$ for the set (\ref{ric3})-(\ref{ric4}) by change of variables $%
\Gamma =\frac{\Phi _{1}}{\Phi _{2}}\,\,$, $g=\frac{\Phi _{3}}{\Phi _{1-}\Phi
_{2}}$.

The super Riccati equations become equivalent to the following linear system
of differential equations
\begin{eqnarray}
D_{\pm }\Psi &=&{\cal A}_{\pm }{\cal \,}\Psi ,  \label{linear} \\
D_{\pm }\Omega &=&{\cal B}_{\pm }{\cal \,}\Omega ,
\label{linear0}
\end{eqnarray}
where
\[
\Psi =\left(
\begin{array}{l}
\Psi _{1} \\
\Psi _{2} \\
\Psi _{3}
\end{array}
\right) ,\qquad \Omega =\left(
\begin{array}{l}
\Phi _{1} \\
\Phi _{2} \\
\Phi _{3}
\end{array}
\right) ,
\]
and ${\cal A}_{\pm }$ and ${\cal B}_{\pm }$ are $3\times 3$ matrices
\begin{eqnarray}
{\cal A}_{+} &=&\sqrt{\frac{1}{2\lambda }}\left(
\begin{array}{lll}
\,\,\,\,\,\,\,\,\,\,0 & \,\,\,\,\,\,\,\,\,\,0 & \,\,\,\cos \Phi \\
\,\,\,\,\,\,\,\,\,\,0 & \,\,\,\,\,\,\,\,\,\,0 & -\sin \Phi \\
2i\cos \Phi & -2i\sin \Phi & \,\,\,\,\,\,\,\,\,0
\end{array}
\right) ,  \nonumber \\
{\cal A}_{-} &=&\frac{1}{2}\,\,\left(
\begin{array}{lll}
\,\,\,\,\,\,\,\,\,\,0 & 2D_{-}\Phi & \sqrt{2\lambda } \\
-2D_{-}\Phi & \,\,\,\,\,\,\,0 & \,\,\,\,\,0 \\
-2i\sqrt{2\lambda } & \,\,\,\,\,\,\,0 & \,\,\,\,\,0
\end{array}
\right) ,  \nonumber \\
{\cal B}_{+} &=&\frac{1}{4}\sqrt{\frac{2}{\lambda }}\,\left(
\begin{array}{lll}
\,\,\,\,\,\,\,\,\,\,0 & \,\,\,\,\,\,\,\,\,\,0 & \,\,\,e^{i\Phi } \\
\,\,\,\,\,\,\,\,\,\,0 & \,\,\,\,\,\,\,\,\,\,0 & -\,e^{-\,i\Phi } \\
4i\,e^{-\,i\Phi } & -4i\,e^{i\Phi } & \,\,\,\,\,\,\,\,\,0
\end{array}
\right) ,  \nonumber \\
{\cal B}_{-} &=&\frac{1}{4}\left(
\begin{array}{lll}
\,\,\,\,\,\,\,\,\,4\,i\,D_{-}\Phi & \,\,\,\,\,\,\,\,\,\,0 & \,\,\sqrt{%
2\lambda }\nonumber \\
\,\,\,\,\,\,\,\,\,\,0 & \,-4\,\,i\,D_{-}\Phi \, & -\,\sqrt{2\lambda }%
\nonumber \\
-4i\sqrt{2\lambda } & 4i\sqrt{2\lambda } & \,\,\,\,\,\,\,\,\,0
\end{array}
\right) .
\end{eqnarray}
The compatibility condition of the linear systems
(\ref{linear})-(\ref{linear0}) in superspace is equivalent to
equation (\ref{sin}). The system (\ref{linear}) is the same as
obtained earlier in \cite{Sc1} but here it is derived
systematically from the B\"{a}cklund transformation of super
sine-Gordon equation. Both linear systems
(\ref{linear})-(\ref{linear0}) are equivalent, related to each
other by a gauge transformation
\begin{equation}
\Psi =G\Omega ,  \label{gau1}
\end{equation}
where
\[
G=\left(
\begin{array}{lll}
-i & i & 0 \\
1 & 1 & 0 \\
0 & 0 & -i
\end{array}
\right) ,
\]
so that
\[
{\cal A}_{\pm }=G{\cal B}_{\pm }G^{-1}.
\]
The matrices ${\cal A}_{\pm }$ and ${\cal B}_{\pm }$ obey the zero curvature
condition in superspace separately, that is
\begin{eqnarray*}
D_{+}{\cal A}_{-}+D_{-}{\cal A}_{+}-\{{\cal A}_{+},{\cal A}_{-}\} &=&0, \\
D_{+}{\cal B}_{-}+D_{-}{\cal B}_{+}-\{{\cal B}_{+},{\cal B}_{-}\} &=&0.
\end{eqnarray*}
The linear systems (\ref{linear})-(\ref{linear0}) can be
reexpressed as
\begin{eqnarray}
\partial _{\pm }\Psi &=&\tilde{{\cal A}}{\cal \,}_{\pm }{\cal \,}\Psi ,
\label{linear2} \\
\partial _{\pm }\Omega &=&\tilde{{\cal B}}{\cal \,}_{\pm }\Omega ,
\label{linear3}
\end{eqnarray}
where the matrices $\tilde{{\cal A}}{\cal \,}_{\pm }$ and $\tilde{{\cal B}}%
{\cal \,}_{\pm }{\cal \,}$are given by
\begin{eqnarray*}
\tilde{{\cal A}}_{+} &=&\left(
\begin{array}{lll}
-\frac{1}{2\lambda }\cos (2\Phi ) & \frac{1}{2\lambda }\sin (2\Phi ) & -%
\frac{i}{2}\sqrt{\frac{2}{\lambda }}\sin (\Phi )D_{+}\Phi \\
\frac{1}{2\lambda }\sin (2\Phi ) & \frac{1}{2\lambda }\cos (2\Phi ) & -\frac{%
i}{2}\sqrt{\frac{2}{\lambda }}\cos (\Phi )D_{+}\Phi \\
\sqrt{\frac{2}{\lambda }}\sin (\Phi )D_{+}\Phi & \sqrt{\frac{2}{\lambda }}%
\cos (\Phi )D_{+}\Phi & -\frac{1}{2\lambda }
\end{array}
\right) , \\
\tilde{{\cal A}}_{-} &=&\left(
\begin{array}{lll}
\frac{\lambda }{2} & \partial _{-}\Phi & 0 \\
-\partial _{-}\Phi & -\frac{\lambda }{2} & \frac{i}{2}\sqrt{2\lambda }%
D_{-}\Phi \\
0 & \sqrt{2\lambda }D_{-}\Phi & \frac{\lambda }{2}
\end{array}
\right) , \\
\tilde{{\cal B}}{\cal \,}_{+} &=&\left(
\begin{array}{lll}
\,\,\,\,\,\,\,\,\,\,\,\,\,\,\,0 & \,\,\,\,\,\,\,\,\,\,\frac{1}{2\lambda }%
\,e^{2i\Phi } & -\frac{1}{4}\sqrt{\frac{2}{\lambda }}e^{i\Phi }D_{+}\Phi \\
\,\,\,\,\,\,\,\,\,\frac{1}{2\lambda }\,e^{-2i\Phi }\,\, & \,\,\,0 & -\frac{1%
}{4}\sqrt{\frac{2}{\lambda }}e^{-i\Phi }D_{+}\Phi \\
i\sqrt{\frac{2}{\lambda }}e^{-i\Phi }D_{+}\Phi & i\sqrt{\frac{2}{\lambda }}%
e^{i\Phi }D_{+}\Phi & \,\,\,\,\,\,\,\,\,\,\,\,\,\,\,\,\,\,\,\,-\,\,\,\frac{1%
}{2\lambda }
\end{array}
\right) , \\
\tilde{{\cal B}}{\cal \,}_{-} &=&\left(
\begin{array}{lll}
\,i\partial _{-}\Phi & \,\,\,\,\,\,\,\,-\,\frac{\lambda }{2} & \frac{\sqrt{%
2\lambda }}{4}D_{-}\Phi \\
\,\,\,\,\,\,\,\,-\,\frac{\lambda }{2} & \,\,\,-i\partial _{-}\Phi & \frac{%
\sqrt{2\lambda }}{4}D_{-}\Phi \\
i\sqrt{2\lambda }D_{-}\Phi & i\sqrt{2\lambda }D_{-}\Phi &
\,\,\,\,\,\,\,\,\,\,\,\,\,\,\,\,\frac{\lambda }{2}\,\,\,\,
\end{array}
\right) .
\end{eqnarray*}
These matrices are also gauge equivalent to each other by the gauge
transformation (\ref{gau1}) i.e
\[
\tilde{{\cal A}}{\cal \,}_{\pm }=G\tilde{{\cal B}}{\cal \,}_{\pm }G^{-1}.
\]
These matrices also obey the following zero curvature condition
\begin{eqnarray*}
\partial _{+}\tilde{{\cal A}}_{-}{\cal -}\partial _{-}\tilde{{\cal A}}_{+}%
{\cal +[}\tilde{{\cal A}}_{-},\tilde{{\cal A}}_{+}{\cal ]} &=&0, \\
\partial _{+}\tilde{{\cal B}}{\cal \,}_{-}{\cal -}\partial _{-}\tilde{{\cal B%
}}{\cal \,}_{+}{\cal +[}\tilde{{\cal B}}{\cal \,}_{-},\tilde{{\cal B}}{\cal %
\,}_{+}{\cal ]} &=&0.
\end{eqnarray*}
The zero curvature condition for the given connection can be reformulated as
a Lax equation with a given Lax operator. The Lax operator associated with
the linear system (\ref{linear2}) is
\[
L=\left(
\begin{array}{lll}
\partial _{-} & \partial _{-}\Phi & 0 \\
-\partial _{-}\Phi & -\partial _{-} & \frac{i}{2}\sqrt{2\lambda }D_{-}\Phi
\\
0 & -\sqrt{2\lambda }D_{-}\Phi & \partial _{-}
\end{array}
\right) .
\]
The Lax equation now immediately follows
\[
\partial _{+}L=[\tilde{{\cal A}_{+}},L].
\]
This equation gives the $x^{+}$-evolution of the spectral problem associated
with the linear eigenvalue equation
\[
L\Psi =\frac{\lambda }{2}\Psi ,
\]
and solves the super sine-Gordon equation in the sense of the inverse
scattering method.

\section{Conclusion}

In summery, we have presented two sets of Riccati equations for
the super sine-Gordon equation. These equations are then used to
derive the super B\"{a}cklund transformation, the linear system
and the linear eigenvalue problem. The linear systems
(\ref{linear})-(\ref{linear0}) obtained by the two super Riccati
equations are shown to be related to each other by a gauge
transformation. By writing the linear system in superspace we have
shown how the zero curvature formulation can be obtained for
fermionic as well as bosonic superfields. This work can be further
extended to investigate the Darboux transformation of the super
sine-Gordon theory. In fact, the Darboux transformation generates
multi-soliton solutions and can be related to the Hirota's
bilinear formalism of the super sine-Gordon equation. These
investigations shall be presented in a separate work.

{\large {\bf Acknowledgment}}

We acknowledge the enabling role of the Higher Education Commission Pakistan
and appreciate its financial support through \textquotedblleft Merit
Scholarship Scheme for PhD studies in Science \& Technology (200
Scholarships)\textquotedblright . We also acknowledge CERN scientific
information Service (publication requests).

\end{document}